\documentclass[aps,prd,preprint,nofootinbib,amsmath,amssymb]{revtex4}

\usepackage{amsmath}
\usepackage{amssymb}
\usepackage{graphicx}
\usepackage{color}
\usepackage{bm}
\usepackage[utf8]{inputenc}
\usepackage{hyperref}
\hypersetup{
  colorlinks=true,
  citecolor=blue,
  linkcolor=blue,
  urlcolor=blue}

\usepackage{soul}

\everymath{\displaystyle}

\usepackage{epsfig}
\usepackage{dcolumn}
\usepackage{float}

\newcommand{\dcp}{\delta_\text{CP}}

\newcommand{\cnv}{\v{C}erenkov}

\begin{document}
\renewcommand{\arraystretch}{1}

\title{Sensitivity of the T2HKK experiment
to the non-standard interaction}

\author{Shinya Fukasawa}
\email[Email Address: ]{fukasawa-shinya@ed.tmu.ac.jp}
\affiliation{Department of Physics, Tokyo Metropolitan University, Hachioji, Tokyo 192-0397, Japan}

\author{Monojit Ghosh}
\email[Email Address: ]{monojit@tmu.ac.jp}
\affiliation{Department of Physics, Tokyo Metropolitan University, Hachioji, Tokyo 192-0397, Japan}

\author{Osamu Yasuda}
\email[Email Address: ]{yasuda@phys.se.tmu.ac.jp}
\affiliation{Department of Physics, Tokyo Metropolitan University, Hachioji, Tokyo 192-0397, Japan}

\begin{abstract}
If the flavor dependent non-standard interactions (NSI)
in neutrino propagation exist, then
the matter effect is modified and
the modification is parametrized by the dimensionless
parameter $\epsilon_{\alpha\beta}~(\alpha,\beta=e, \mu, \tau)$.
In this paper we discuss the sensitivity
of the T2HKK experiment, whose possibility
is now seriously discussed as a future extension of
the T2K experiment, to such NSI.
On the assumption that
$\epsilon_{\alpha\mu}=0~(\alpha=e, \mu, \tau)$
and $\epsilon_{\tau\tau}=|\epsilon_{e\tau}|/(1+\epsilon_{ee})$,
which are satisfied by other experiments
to a good approximation, we find that,
among the possible off-axis flux configurations of
1.3$^\circ$, $1.5^\circ$, $2.0^\circ$ and 2.5$^\circ$,
the case of the off-axis angle 1.3$^\circ$
gives the highest sensitivity to $\epsilon_{ee}$
and $|\epsilon_{e\tau}|$. 
Our results show that the $1.3^\circ$ off-axis configuration can exclude NSI for
$|\epsilon_{ee}|\gtrsim 1$
or $|\epsilon_{e\tau}|\gtrsim 0.2$
at 3$\sigma$. 
We also find that in the presence of NSI,
T2HKK (for the off-axis angle 1.3$^\circ$)
has better sensitivity to the two CP phases ($\dcp$ and arg($\epsilon_{e \tau}$))
than DUNE.
This is because of the synergy
between the two detectors i.e., one at Kamioka and one at Korea.
T2HKK has better sensitivity
to the CP phases than
the atmospheric neutrino experiment
at Hyperkamiokande in inverted hierarchy,
but in normal hierarchy 
the atmospheric neutrino experiment has the best sensitivity
to the CP phases.
\end{abstract}

\maketitle


\section{Introduction}
It has been established by the successful experiments
in the past that neutrinos have masses and mixings\,\cite{Olive:2016xmw}.
The three mixing angles $\theta_{12}$, $\theta_{13}$, $\theta_{23}$
and two mass-squared differences $\Delta m^2_{31}$, $\Delta m^2_{21}$
in the standard three flavor neutrino oscillation framework
are measured as:
$(\Delta m^2_{21}, \sin^22\theta_{12}) \simeq (7.5\times 10^{-5}$eV$^2$, $0.86)$,
$(|\Delta m^2_{31}|, \sin^22\theta_{23}) \simeq (2.4\times 10^{-3}$eV$^2$, $1.0)$,
$\sin^22\theta_{13}\simeq 0.09$.
The remaining unknowns are the value of the Dirac CP phase $\dcp$,
the sign of $\Delta m^2_{31}$ (the mass hierarchy i.e., normal or inverted) and the octant
of $\theta_{23}$ (the sign of $\pi/4-\theta_{23}$ i.e., lower or higher).
It is expected that these unknowns will be
determined by the future neutrino oscillation experiments,
particularly the accelerator based long-baseline neutrino
experiments\,\cite{Abe:2014oxa,Acciarri:2015uup}.
These experiments in the future can not only
measure the oscillation parameters in the standard
three flavor mixing scenario but also probe 
the new physics by looking at the deviation from
the standard three flavor neutrino mixing framework.

Flavor-dependent neutral current neutrino Non-Standard Interactions
(NSI)\,\cite{Wolfenstein:1977ue,Guzzo:1991hi,Roulet:1991sm} have been
studied as one of the new physics candidates which can be searched at
the future neutrino experiments \cite{Ohlsson:2012kf,Miranda:2015dra}.
In the presence of these NSI the
neutrino propagation feels the extra contribution to the matter effect
and hence long-baseline experiments with a longer baseline length $L$
(typically $L \gtrsim$1000 km) and the atmospheric neutrino experiments
are expected to have
sensitivity to the neutral current NSI. Recent studies of neutral current NSI in long-baseline and atmospheric neutrino experiments can be found in 
Ref. \cite{
Friedland:2012tq,
Adhikari:2012vc,
Esmaili:2013fva,
Choubey:2014iia,
Chatterjee:2014gxa,
Masud:2015xva,
deGouvea:2015ndi,
Rahman:2015vqa,
Fukasawa:2015jaa,
Choubey:2015xha,
Coloma:2015kiu,
Liao:2016hsa,
Soumya:2016enw,
Blennow:2016etl,
Forero:2016cmb,
Huitu:2016bmb,
Bakhti:2016prn,
Masud:2016bvp,
Coloma:2016gei,
Masud:2016gcl,
Agarwalla:2016fkh,
Ge:2016dlx,
Fukasawa:2016nwn,
Liao:2016bgf,
Fukasawa:2016gvm,
Blennow:2016jkn
}. 

The possibility of a second detector in Korea for the T2K \cite{Abe:2014tzr}
experiment was discussed in the past \cite{Kim:2000kias,Hagiwara:2004iq,Ishitsuka:2005qi,Hagiwara:2005pe,Hagiwara:2006vn,Kajita:2006bt,Barger:2007jq,Kimura:2007mu,
Ribeiro:2007jq,Huber:2007em,Hagiwara:2009bb,Oki:2010uc,Hagiwara:2011kw,Hagiwara:2012mg,Dufour:2012zr,Hagiwara:2016qtb}.
Recently there has been a renewed interest in the idea of
placing the second detector in Korea as a part of the T2HK plan \cite{Abe:2014oxa}, and
the plan with the second detector in Korea is now called
the T2HKK project \cite{Nakaya:2016nufact}.
The original plan of the T2HK project is to build a large tank of water \cnv\ detector 
at the Kamioka site. 
Under the T2HKK project, there will be two tanks of equal volume 
instead of
building a single tank and then one of the tanks will be built in Korea.
Depending on the location of the second detector
in Korea, one has different options for the flux in terms of the off-axis angle.
According to the HK collaboration, there are various flux options between $1^\circ$ to $3^\circ$ off-axis configurations are under consideration at present \cite{T2HKK}. 
In the T2HKK project, there are some discussions
on which location is the most advantageous
from the physics point of view.
In this paper for the first time we study the sensitivity of T2HKK
to NSI and discuss the result of optimization for the
NSI parameters $\epsilon_{ee}$ and $|\epsilon_{e\tau}|$ with respect to the different flux options.
We also compare its sensitivity with that of DUNE \cite{Acciarri:2015uup} and the
atmospheric neutrino experiment at Hyperkamiokande (HK)\,\cite{Abe:2011ts}.
While a similar analysis was done in the past\,\cite{Oki:2010uc},
the new points in the present paper are
the optimization with respect to the location,
which can be expressed in terms of the off-axis angle,
and the comparison of the sensitivity with DUNE and the atmospheric neutrino at 
HK.

This paper is organized as follows. In Section \ref{preliminaries}, we
describe the constraints on NSI in propagation. In Section \ref{sensitivity}, 
we study the sensitivity of the T2HKK experiment to NSI and compare our results with DUNE and HK. We will also compare our results with the T2HK configuration. In Section
\ref{conclusion}, we draw our conclusions.  

\section{Preliminaries}
\label{preliminaries}

\subsection{Nonstandard interactions}
\label{nsi_intro}

Let us start with the effective flavor-dependent neutral current
neutrino non-standard interactions in propagation given by
\begin{eqnarray}
{\cal L}_{\mbox{\rm\scriptsize eff}}^{\mbox{\tiny{\rm NSI}}} 
=-2\sqrt{2}\, \epsilon_{\alpha\beta}^{ff'P} G_F
\left(\overline{\nu}_{\alpha L} \gamma_\mu \nu_{\beta L}\right)\,
\left(\overline{f}_P \gamma^\mu f_P'\right),
\label{NSIop}
\end{eqnarray}
where $f_P$ and $f_P'$ stand for fermions with chirality $P$ and
$\epsilon_{\alpha\beta}^{ff'P}$ is a dimensionless constant
which is normalized by the Fermi coupling constant $G_F$.
The presence of NSI in Eq.\.(\ref{NSIop}) modifies the MSW potential in the
flavor basis from
\begin{eqnarray}
 \sqrt{2} G_F N_e\left(
\begin{array}{ccc}
1 &  0& 0\\
0 &0  &0\\
0 &0 & 0
\end{array}
\right)
\end{eqnarray}
to
\begin{eqnarray}
{\cal A} \equiv
\sqrt{2} G_F N_e \left(
\begin{array}{ccc}
1+ \epsilon_{ee} & \epsilon_{e\mu} & \epsilon_{e\tau}\\
\epsilon_{\mu e} & \epsilon_{\mu\mu} & \epsilon_{\mu\tau}\\
\epsilon_{\tau e} & \epsilon_{\tau\mu} & \epsilon_{\tau\tau}
\end{array}
\right),
\label{matter-np}
\end{eqnarray}
where $\epsilon_{\alpha\beta}$ is defined by
\begin{equation}
\epsilon_{\alpha\beta}\equiv\sum_{f=e,u,d}\frac{N_f}{N_e}\epsilon_{\alpha\beta}^{f}\,.
\end{equation}
$N_f~(f=e, u, d)$ stands for number densities of fermions $f$.
Here we defined the NSI parameters as
$\epsilon_{\alpha\beta}^{fP}\equiv\epsilon_{\alpha\beta}^{ffP}$ and
$\epsilon_{\alpha\beta}^{f}\equiv\epsilon_{\alpha\beta}^{fL}+\epsilon_{\alpha\beta}^{fR}$.
In the three flavor
neutrino oscillation framework with NSI, the neutrino evolution is
given by the Schrodinger equation:
\begin{eqnarray}
i {d \over dx} \left( \begin{array}{c} \nu_e(x) \\ \nu_{\mu}(x) \\ 
\nu_{\tau}(x)
\end{array} \right)  = 
\left[  U {\rm diag} \left(0, \Delta E_{21}, \Delta E_{31}
\right)  U^{-1} 
+ {\cal A}\right]
\left( \begin{array}{c}
\nu_e(x) \\ \nu_{\mu}(x) \\ \nu_{\tau}(x)
\end{array} \right)\,,
\label{eqn:sch}
\end{eqnarray}
where
$U$ is the leptonic mixing matrix defined by
\begin{eqnarray}
U&\equiv&\left(
\begin{array}{ccc}
c_{12}c_{13} & s_{12}c_{13} &  s_{13}e^{-i\dcp}\cr
-s_{12}c_{23}-c_{12}s_{23}s_{13}e^{i\dcp} & 
c_{12}c_{23}-s_{12}s_{23}s_{13}e^{i\dcp} & s_{23}c_{13}\cr
s_{12}s_{23}-c_{12}c_{23}s_{13}e^{i\dcp} & 
-c_{12}s_{23}-s_{12}c_{23}s_{13}e^{i\dcp} & c_{23}c_{13}
\end{array}
\right),
\label{eqn:mns3}
\end{eqnarray}
and $\Delta E_{jk}\equiv\Delta m_{jk}^2/2E\equiv (m_j^2-m_k^2)/2E$,
$c_{jk}\equiv\cos\theta_{jk}$, $s_{jk}\equiv\sin\theta_{jk}$.

As far as the neutrino oscillation on the Earth
is concerned, we have the following limits
on $\epsilon_{\alpha\beta}$ from the compilation of 
various neutrino data at 90\% C.L:\cite{Davidson:2003ha,Biggio:2009nt}\footnote{
See Ref.\,\cite{Grossman:1995wx} for the constraints
on NSI at production and detection.}
\begin{eqnarray}
\left(
\begin{array}{ccc}
|\epsilon_{ee}| < 4\times 10^0 & \quad|\epsilon_{e\mu}| < 3\times 10^{-1}
& \quad|\epsilon_{e\tau}| < 3\times 10^{0\ }\\
& \quad |\epsilon_{\mu\mu}| < 7\times 10^{-2}
& \quad|\epsilon_{\mu\tau}| < 3\times 10^{-1}\\
& & \quad|\epsilon_{\tau\tau}| < 2\times 10^{1\ }
\end{array}
\right).
\label{epsilon-m}
\end{eqnarray}
It was pointed out in Refs.\,\cite{Friedland:2004ah,Friedland:2005vy}
that the high-energy atmospheric neutrino data, where the matter
effects are dominant, are consistent with NSI only when the following
equality is approximately satisfied:
\begin{eqnarray}
\epsilon_{\tau\tau}=\frac{|\epsilon_{e\tau}|^2}{1+\epsilon_{ee}}\,.
\label{eq:ansatz_a}
\end{eqnarray}
In this paper we assume the relation (\ref{eq:ansatz_a}) exactly.
It may seem that the condition (\ref{eq:ansatz_a}) forces
$|\epsilon_{e\tau}|$ to be smaller than it should be,
near the region $|1+\epsilon_{ee}| \ll 1$.  However,
Eq.\,(\ref{eq:ansatz_a}) turns out to be a reasonable condition
even in the region $|1+\epsilon_{ee}| \ll 1$ because of the
following arguments.
The constraint from the high energy atmospheric
neutrino data is that the smaller eigenvalue in the
matter potential (\ref{matter-np}) with $\epsilon_{\alpha\mu}=0$
should be smaller than $0.2\times\sqrt{2} G_F N_e$ (See Eq.(13) in
Ref.\,\cite{Fukasawa:2016nwn}).  For $|1+\epsilon_{ee}| \ll 1$,
this implies 
$\sqrt{4|\epsilon_{e\tau}|^2+\epsilon_{\tau\tau}^2}-\epsilon_{\tau\tau}<0.4$.  
This condition in principle allows us to take a large
value of $|\epsilon_{e\tau}|$.  However, we have checked explicitly
that such a large value of $|\epsilon_{e\tau}|$ gives a very bad fit
to the HK atmospheric neutrino data assuming the standard scenario,
and therefore $|\epsilon_{e\tau}|\ll 1$
should be satisfied near the region $|1+\epsilon_{ee}| \ll 1$.\footnote{
At present we have only the very weak bound on $|\epsilon_{e\tau}|$.
If we find $|1+\epsilon_{ee}|\ll 1$ experimentally in the future,
however, then by combining $|1+\epsilon_{ee}|\ll 1$ and
the high energy atmospheric neutrino data,
we will obtain the very strong bound on $|\epsilon_{e\tau}|$.
}
This result is consistent with the discussion in Ref.\,\cite{Oki:2010uc}
based on an analytic formula on the disappearance probability.
Namely, the high energy atmospheric neutrino data is perfectly
consistent with its behavior $1-P(\nu_\mu\to\nu_\mu)\propto 1/E^2$
inferred from the standard oscillation scenario, while
in the presence of NSI with $\epsilon_{\mu\alpha}=0$,
it has the behavior $1-P(\nu_\mu\to\nu_\mu)\sim c_1/E+ O(1/E^2)$
where $c_1$ satisfies $c_1 \propto \epsilon_{\tau\tau}-|\epsilon_{e\tau}|^2/(1+\epsilon_{ee})$
in the case of $|\epsilon_{\tau\tau}-|\epsilon_{e\tau}|^2/(1+\epsilon_{ee})|\ll 1$
(See Eq.(9) in Ref.\,\cite{Oki:2010uc}).
Therefore, 
the condition (\ref{eq:ansatz_a}) is a good approximation
also in the region $|1+\epsilon_{ee}| \ll 1$, and the 
ansatz (\ref{eq:ansatz_a}) is justified.

If Eq.\,(\ref{eq:ansatz_a}) is satisfied, then
$\epsilon_{\tau\tau}$ can be eliminated. 
Furthermore, we have
\begin{eqnarray}
\left|\frac{\epsilon_{e\tau}}{1+\epsilon_{ee}}\right|
\lesssim 0.8\quad\mbox{\rm at}~3\sigma \mbox{\rm C.L.,}\,.
\label{tanb}
\end{eqnarray}
from the atmospheric neutrino data of Superkamiokande \cite{Fukasawa:2015jaa}.

From the above two constraints (\ref{epsilon-m})
and (\ref{eq:ansatz_a}), the following ansatz
is a good approximation to analyze the sensitivity to NSI:
\begin{eqnarray}
{\cal A} = \sqrt{2} G_F N_e\left(
\begin{array}{ccc}
1+\epsilon_{ee}&0&\epsilon_{e\tau}\cr
0&0&0\cr
\epsilon_{e\tau}^\ast&0&|\epsilon_{e\tau}|^2/(1+\epsilon_{ee})
\end{array}\right)\,.
\label{ansatz}
\end{eqnarray}
The allowed region in the ($\epsilon_{ee}$,$|\epsilon_{e\tau}|$)
plane at 90\% C.L., is given by the following:
\begin{eqnarray}
-4 \lesssim \epsilon_{ee} \lesssim 4,~
|\epsilon_{e\tau}|\lesssim 3,~
\left|\frac{\epsilon_{e\tau}}{1+\epsilon_{ee}}\right|
\lesssim 0.6\,.
\label{ett}
\end{eqnarray}

\subsection{The T2HKK experiment
\label{t2kk}}

\begin{figure*}
\hspace{-30pt}
\includegraphics[scale=1.1]{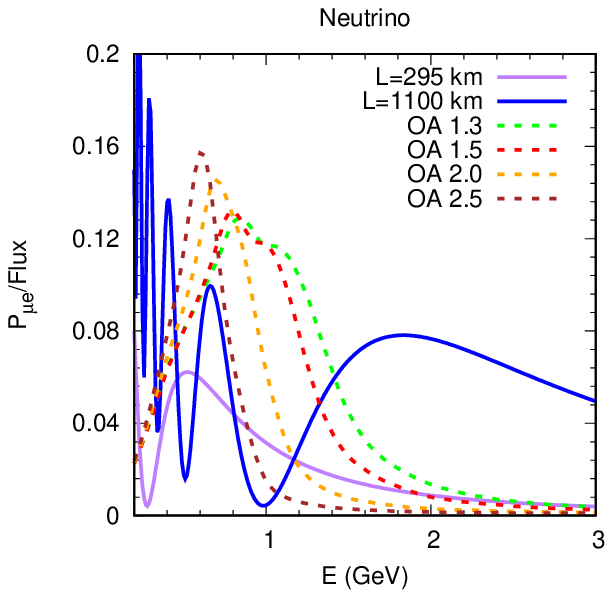}
\hspace{-80pt}
\includegraphics[scale=1.1]{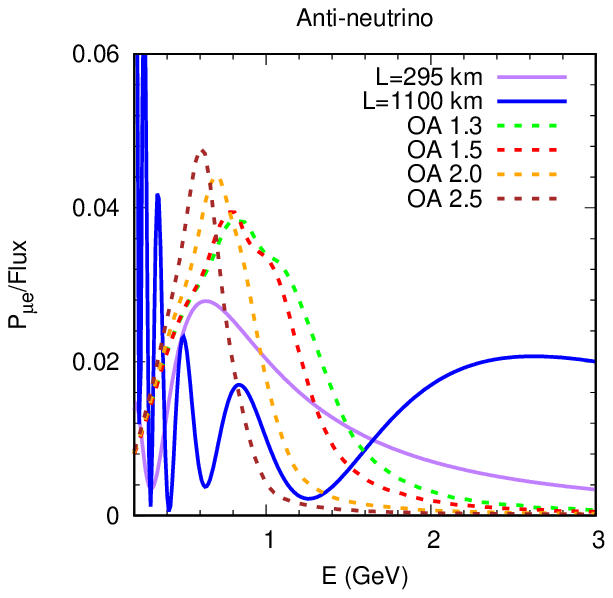}
\caption{The flux (dashed curves) at different off-axis
angles and the appearance oscillation probabilities
(solid curves) at Kamioka ($L$=295 km) and in Korea
($L$=1100 km) in the
standard oscillation scenario in normal hierarchy.
The left (right) panel is for neutrinos (antineutrinos).
The baseline length $L$=1088 km at an angle 1.3$^\circ$
is slightly different from $L$=1100 km, but the
difference between the
oscillation probabilities at $L$=1088 km and
at $L$=1100 km is invisibly small.}
\label{fig:fig1}
\end{figure*}

The T2HKK experiment
is a proposal for the future extension
of the T2K experiment.
In this proposal, a water
\cnv\,\,detector
is placed not only in Kamioka (at a baseline length $L$ = 295 km) but also
in Korea (at $L\simeq$ 1100 km), whereas the power of the beam
at J-PARC in Tokai Village is upgraded to 1.3 MW.
As in the T2K experiment, it is assumed that
T2HKK uses an off-axis beam at a 2.5$^\circ$
angle between the directions
of the decaying charged pions and neutrinos,
and the neutrino energy spectrum has a peak approximately
at 0.6 GeV.
This off-axis beam at an angle 2.5$^\circ$ reaches Korea
and the corresponding off-axis angle on the surface
in Korea ranges from 1.3$^\circ$ to 2.5$^\circ$
with the baseline 1088 km (for 1.3$^\circ$)
to 1100 km (for 1.5$^\circ$, 2.0$^\circ$ and 2.5$^\circ$), depending on
the location of the detector in Korea \footnote{The other flux options which are also under consideration are: $1.8^\circ$, $1.9^\circ$ and $2.2^\circ$ \cite{T2HKK} which are not considered in this work.}.
The flux and the appearance oscillation
probabilities for neutrinos and antineutrinos
at various off-axis angles in normal hierarchy are shown in
Fig.\,\ref{fig:fig1}. 
The label in the $y$ axis corresponds to the value of $P_{\mu e}$ whereas the the unit of the fluxes are arbitrary.
As we can see from Fig.\,\ref{fig:fig1},
the first oscillation maximum occurs at
$E\simeq$ 1.8 (2.6) GeV, whereas the second one
appears at $E\simeq$ 0.7 (0.8) GeV for
neutrinos (antineutrinos). From Fig. \ref{fig:fig1} we observe the following:
\begin{itemize}
 \item  Among the different off-axis fluxes, the flux corresponding to the lowest off-axis angle (i.e., OA 1.3) peaks at 0.8 GeV  and the flux at the highest off-axis angle (i.e., OA 2.5) peaks at 0.6 GeV.
 \item  The relative heights of the peak of the fluxes is maximum for OA 2.5 and minimum for OA 1.3. 
 \item  The off-axis fluxes corresponding $2.5^\circ$ and $2.0^\circ$ can mainly probe the physics at the second oscillation maxima for $L=1100$ km while the 
 off-axis fluxes corresponding to $1.3^\circ$ and $1.5^\circ$ can also cover a part of the first oscillation maxima for the Korean baseline.
\end{itemize}

\section{Sensitivity of T2HKK to $\epsilon_{ee}$ and $|\epsilon_{e\tau}|$
\label{sensitivity}}

In this section we discuss the sensitivity of the T2HKK
experiment to the non-standard interaction in propagation 
with the ansatz (\ref{ansatz}). For comparison, we also study
sensitivity of the DUNE \cite{Acciarri:2015uup} and
the atmospheric neutrino experiment at HK\,\cite{Abe:2011ts}.
Since $\epsilon_{\tau\tau}$ is expressed in terms of
$\epsilon_{ee}$ and $|\epsilon_{e\tau}|$,
the only new degrees of freedom are
$\epsilon_{ee}$, $|\epsilon_{e\tau}|$
and $\mbox{\rm arg}(\epsilon_{e\tau})$.
First of all,
in sect.\,\ref{discrimination},
assuming that the Nature is described
by the standard three-flavor scheme,
we discuss the bounds on $\epsilon_{ee}$
and $|\epsilon_{e\tau}|$.
In our analysis
we assume that the true numbers of events
are those of the standard three-flavor scenario,
and the test numbers of events
are those with NSI.
We discuss the region of
the $(\epsilon_{ee}, |\epsilon_{e\tau}|)$ plane
in which T2HKK can exclude the hypothesis with NSI.
Secondly, in sect.\ref{cp},
assuming that NSI exists,
we consider whether
the two complex phases
$\dcp$ and $\mbox{\rm arg}(\epsilon_{e\tau})$
can be determined separately.

The neutrino flux of the T2HKK experiment in Korea
is taken from Ref.\,\cite{Hyper-Kamiokande:2016}.
 To calculate the event rates for the T2HKK setup we proceed in the following way. 
 First we have matched the number of events corresponding to the T2HK setup as given in Ref. \cite{Abe:2014oxa} taking the $2.5^\circ$ off-axis flux.
The detector volume in this case is 560 kt.
Then we scale these number of events for the Korean baseline corresponding to different off-axis configurations. For T2HKK project we have taken 280 kt detector both at Kamioka and Korea.
Note that as we have taken the backgrounds corresponding to the T2HK setup and scale them down for the Korean baseline, the neutral current $\pi^0$
backgrounds at the high energies are ignored and thus our results of T2HKK may be optimistic. For T2HKK setup we have taken a total integrated beam-power of $15.6 \times 10^{21}$
pot (protons on target) with $10^{21}$ pot/year. Thus it corresponds to 15.6 years running of the beam. For T2HKK we have taken an overall systematic error of 3.3\% for both appearance and disappearance channel in neutrino mode and 
6.2\% (4.5\%) for appearance (disappearance) channel in antineutrino mode. 
The systematic error is the same for both the signal and the background 
\footnote{ Note that in our work we followed the configuration of T2HK as given in \cite{Abe:2014oxa}. 
According to the T2HKK report \cite{T2HKK} (which appeared on the same day as our paper appeared in arXiv), the total detector volume is around
380 kt which will be split into 190 kt for each Kamioka and Korea. The total exposure in this report is $2.7 \times 10^{22} $ pot with 10 years of running.}.  
For DUNE we have taken a flux of beam-power 1.2 MW with  $10^{21}$ pot/year and 34 kt liquid argon detector. 
In our analysis we have considered a 10 years running of DUNE unless otherwise mentioned.
The number of events are taken from Ref. \cite{Acciarri:2015uup}. 
The systematic error for DUNE is 2\% (10\%) for appearance channel and 5\% (15\%) for disappearance channel corresponding to the signal (the background). The systematic errors in neutrino
and antineutrino mode are the same for DUNE.
The simulations of T2HKK and DUNE
have been performed with
the softwares GLoBES\,\cite{Huber:2004ka,Huber:2007ji} and MonteCUBES \cite{Blennow:2009pk}.

Assuming the operation with $\nu$:$\bar{\nu}$ = 1:1,
as well as the oscillation parameters
$\theta_{23}=\pi/4$, $\dcp=-\pi/2$ with normal hierarchy,
the expected numbers of appearance events in Korea, 
are shown in Table \ref{tab:korea},
while those at Kamioka are
3219 neutrinos and 420 antineutrinos.
The expected numbers of appearance events at DUNE
are 1897 neutrinos and 229 antineutrinos.
Thus we understand that among all the off-axis configurations, the number of events are maximum for $1.3^\circ$. This is because the $1.3^\circ$ off-axis configuration covers the major portion of the
first oscillation maxima where the appearance channel probability has a significant contribution (c.f Fig. \ref{fig:fig1}). From the above discussion it is also clear that the number of events at the Kamioka detector
is almost 1.8 times that of the number of events for DUNE.
Simulation of the atmospheric neutrino
at HK is done with
the codes which were
used in Refs.\,\cite{Foot:1998iw,Yasuda:1998mh,Yasuda:2000de,Fukasawa:2015jaa}
and is described in detail in Ref.\,\cite{Fukasawa:2015jaa}.
We assume here the data size from the HK atmospheric neutrino
experiment for 15 years with 560 kt fiducial volume.

\begin{table}
\begin{center}
\begin{tabular}{|c|c|c|c|c|}
\hline
Off-axis degree   & 1.3$^\circ$ & 1.5$^\circ$  & 2.0$^\circ$ & 2.5$^\circ$\\          
\hline
Neutrinos &    515   & 438    &  368  & 309    \\
Antineutrinos  & 39  & 34 & 25 & 17\\
\hline
\end{tabular}
\end{center}
\caption{The numbers of appearance events for neutrino and antineutrinos
expected at the second detector in Korea.
$\theta_{23}=\pi/4$, $\delta=-\pi/2$ with
normal hierarchy is assumed.
}
\label{tab:korea} 
\end{table}

\subsection{Bounds on $\epsilon_{ee}$ and $|\epsilon_{e\tau}|$
\label{discrimination}}

\begin{figure*}
\hspace{-20pt}
\includegraphics[scale=0.8]{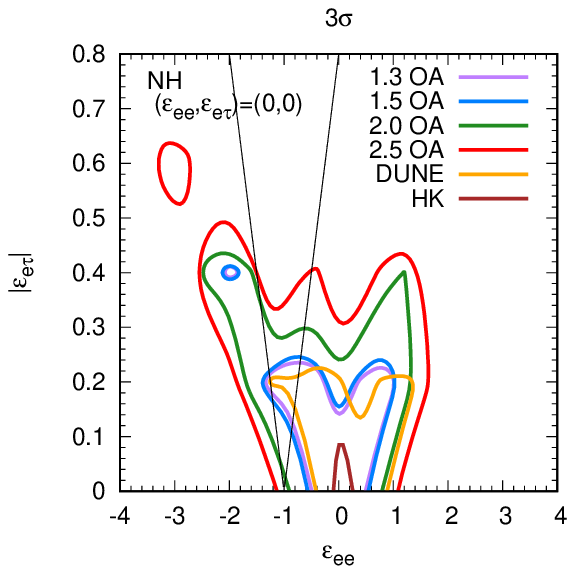}
\hspace{-80pt}
\includegraphics[scale=0.8]{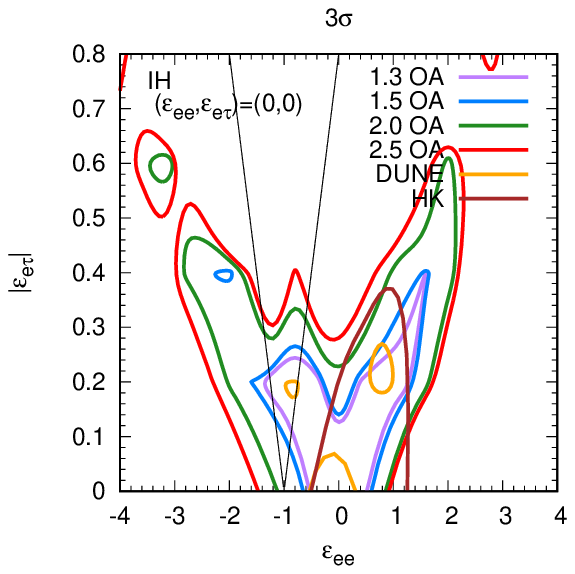}
\hspace{-60pt}
\includegraphics[scale=0.8]{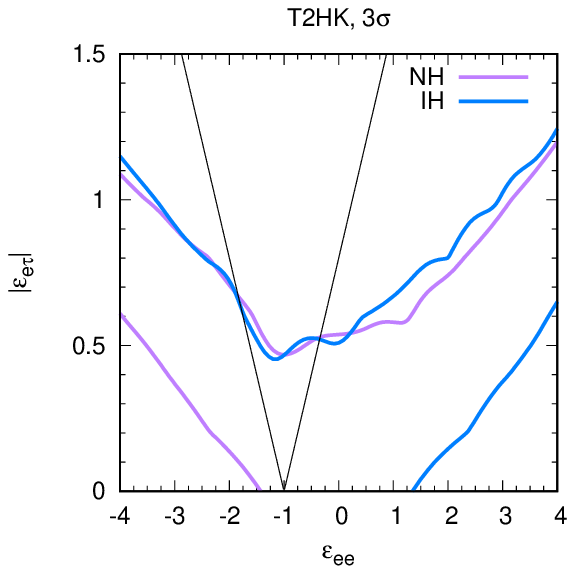}
\caption{
The excluded region in the
($\epsilon_{ee}$, $|\epsilon_{e\tau}|$) plane.
The hypothesis with NSI is excluded at 3$\sigma$
outside each curve.  The thin solid diagonal straight line
stands for the bound $|\tan\beta|\equiv
|\epsilon_{e\tau}/(1+\epsilon_{ee})|\lesssim 0.8$\,\cite{Fukasawa:2015jaa}
at 3$\sigma$ from the current atmospheric
data by Superkamiokande.  Upper left pane:
Normal mass hierarchy.  Upper right panel:
Inverted  mass hierarchy.  Lower panel:
The bounds from T2HK
with the detector of volume 560 kt at Kamioka only.
}
\label{fig:fig2}
\end{figure*}

Firstly, let us discuss the case of the
region ($\epsilon_{ee}$, $|\epsilon_{e\tau}|$), in which
we can test the difference between NSI with ansatz (\ref{ansatz})
and the standard three-flavor scheme.
Here, we take the best-fit values
for most of the standard oscillation parameters
as the reference values:\footnote{
The oscillation parameters with
bars (without bars) stands for the true
(test) value throughout this paper.}
\begin{eqnarray}
\sin^2 (2\bar{\theta}_{12}) &=& 0.87 
\nonumber\\
\sin^2 (2\bar{\theta}_{23}) &=& 1.0
\nonumber\\
\sin^2 (2\bar{\theta}_{13}) &=& 0.085
\nonumber\\
\Delta \bar{m}_{21}^2 &=& 7.9 \times 10^{-5} {\rm eV}^2
\nonumber\\
\Delta \bar{m}_{32}^2 &=& 2.4 \times 10^{-3} {\rm eV}^2
\nonumber\\
\bar{\delta}_{\mbox{\rm\scriptsize CP}} &=& -90^\circ
\label{central-values}
\end{eqnarray}
For the parameters $\theta_{12}$, $\theta_{13}$, $\Delta m_{21}^2$ and $\Delta m_{32}^2$ our choice of true parameters are consistent with the best-fit values as obtained by the global
analysis of the world neutrino data \cite{Forero:2014bxa,Esteban:2016qun,Capozzi:2013csa}. The status of $\theta_{23}$ at this moment is quite intriguing. The latest T2K data favours maximal mixing \cite{t2k}
whereas the NO$\nu$A data disfavours maximal mixing at $2.5 \sigma$ \cite{nova}. Thus one needs more data from both the experiments to resolve this issue. For our work we have taken $\theta_{23}$ to be maximal in
the true spectrum and marginalized from $40^\circ$ to $50^\circ$ in the test spectrum. 
Regarding $\dcp$ both the experiments obtain a best-fit value of $-90^\circ$ which we also take as true value in our analysis. We have marginalized $\dcp$ in the test spectrum 
from $-180^\circ$ to $180^\circ$. For the NSI parameters we have taken $\bar{\epsilon}_{ee} = |\bar{\epsilon}_{e \tau}| = \mbox{\rm arg}(\bar{\epsilon}_{e\tau}) = 0$. In the test spectrum we have marginalized over 
$\mbox{\rm arg}(\epsilon_{e\tau})$ from $-180^\circ$ to $180^\circ$.
The results are shown in Fig.\,\ref{fig:fig2},
where the curves are drawn at 3$\sigma$
($\Delta\chi^2=11.83$ for 2 degrees of freedom).
NSI with the ansatz (\ref{ansatz}) can be
distinguished from the standard three-flavor scheme
outside the curves.  For comparison, we have also
showed the excluded
regions by the long-baseline experiment DUNE , the atmospheric neutrino experiment
HK and T2HK
with the detector of volume 560 kt at Kamioka only.
From Fig.\,\ref{fig:fig2},
we observe that the case at off-axis angle 1.3$^\circ$
has the highest sensitivity to 
($\epsilon_{ee}$, $|\epsilon_{e\tau}|$).
This is because (i) the number of events
is the largest at off-axis angle 1.3$^\circ$,
as we can see from Table \ref{tab:korea} and (ii) due to the relatively broad band nature of the flux, the $1.3^\circ$ off axis configuration covers the wider energy range in the probability 
spectrum among the other off-axis configurations.  
The sensitivity at the $1.5^\circ$ off-axis is similar as that of $1.3^\circ$ while the sensitivities at $2.0^\circ$ and $2.5^\circ$ are worse than sensitivities at $1.3^\circ$ and $1.5^\circ$. 
If we compare the sensitivity of the T2HKK with T2HK, then we find that T2HKK is far more powerful than T2HK in terms of constraining the value of the NSI parameters. 
These results are true for both normal and inverted hierarchies.
The sensitivity of DUNE to NSI is comparable
to T2HKK at 1.3$^\circ$ for normal hierarchy and better in inverted hierarchy. The sensitivity of HK atmospheric
neutrino experiment is the highest for both the hierarchies. In Table Table.\,\ref{tab:eeet} we have given the $90\%$ C.L., as well as $3 \sigma$ bounds on $\epsilon_{ee}$ and $|\epsilon_{e\tau}|$ for the
different experimental setups which are considered in our analysis. From the table we see that the sensitivity of the $1.3^\circ$ configuration in constraining ($\epsilon_{ee}$, $|\epsilon_{e\tau}|$) 
is one order of magnitude higher than the configuration of $2.5^\circ$.

\begin{table}
\begin{center}
\begin{tabular}{|c|c|c|}
\hline
Experiment   & $\epsilon_{ee}$ $90\%$ C.L ($3 \sigma$) & $|\epsilon_{e\tau}|$ $90\%$ C.L ($3 \sigma$)\\ 
\hline
T2HK & -4 to +4 (-4 to +4) & $<$ 0.9 ($<$ 1.1)\\
T2HKK(OA1.3$^\circ$) & -0.2 to 0.2 (-1.4 to 1.1) & $<$ 0.02 ($<$ 0.24) \\
T2HKK(OA1.5$^\circ$)& -0.2 to 0.2 (-1.4 to 1.1) & $<$ 0.02 ($<$ 0.24)\\
T2HKK(OA2.0$^\circ$) & -1.2 to 0.6 (-3.5 to 1.4) & $<$ 0.03 ($<$ 0.44 )\\
T2HKK(OA2.5$^\circ$)& -1.4 to 1.0 (-3.5 to 1.8)& $<$ 0.2 ($<$ 0.5)\\
DUNE & -0.1 to 0.4 (-1.2 to 1.4) & $<$ 0.04 ($<$ 0.23)\\
atm(HK) & -0.05 to 0.1 (-0.1 to 0.2) &$<$ 0.035 ($<$ 0.1) \\          
\hline
\end{tabular}
\end{center}
\caption{The bounds on $\epsilon_{ee}$ and $|\epsilon_{e\tau}|$
at 90\% C.L., ($3 \sigma$) by each experiment in the case of normal hierarchy.
}
\label{tab:eeet} 
\end{table}

\begin{figure*}
\hspace{-30pt}
\includegraphics[scale=0.8]{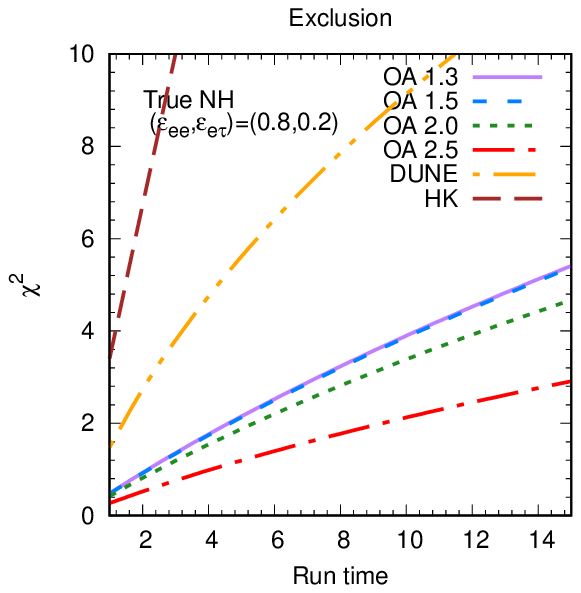}
\hspace{-75pt}
\includegraphics[scale=0.8]{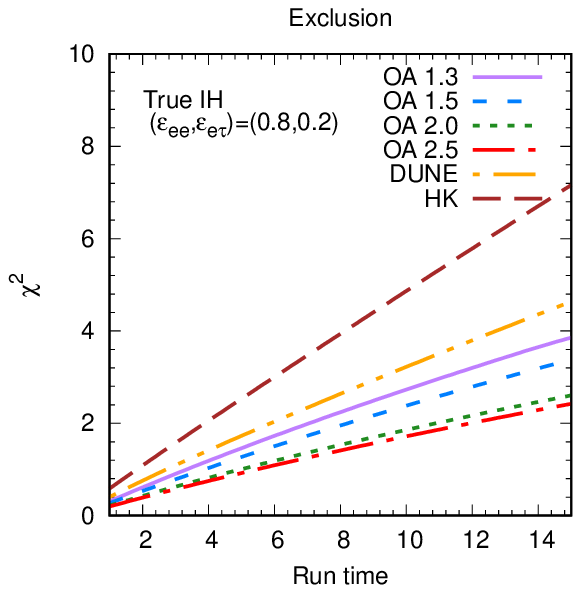}
\hspace{-54pt}
\includegraphics[scale=0.8]{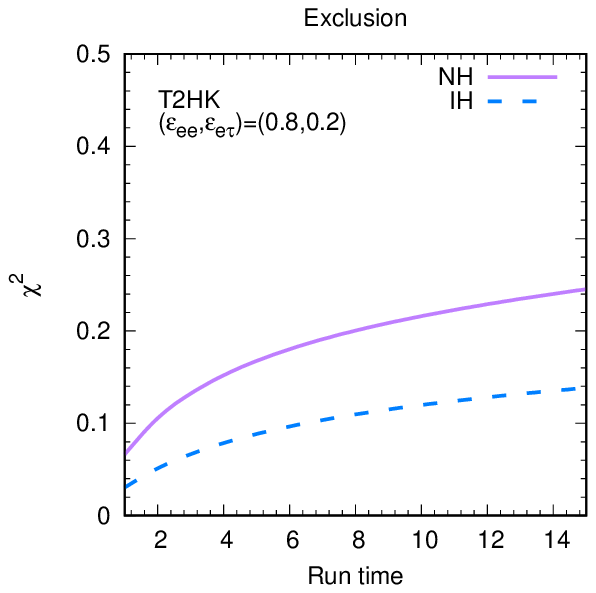}
\caption{$\chi^2$ to exclude
($\epsilon_{ee}$, $|\epsilon_{e\tau}|$) = (0.8, 0.2)
as a function of the running time.
  Upper right panel:
Inverted  mass hierarchy.  Lower panel:
The bounds from T2HK
with the detector at Kamioka only.
} 
\label{fig:fig3}
\end{figure*}

In Fig.\,\ref{fig:fig3}, $\chi^2$ to exclude
a particular choice 
($\epsilon_{ee}$, $|\epsilon_{e\tau}|$) = (0.8, 0.2)
is plotted as a function of the running time.
Here for comparison we have extended the DUNE runtime to 15 years.
From the figures we see that the sensitivity in normal hierarchy is better than inverted hierarchy.
This is because in inverted hierarchy the MSW effect enhances the antineutrino probabilities and the cross section of the antineutrinos are almost one third of the neutrinos. Thus the number 
of events in the IH are less as compared to the normal hierarchy.
Similar as that of Fig. \ref{fig:fig2}, the sensitivity of $1.5^\circ$ is comparable with $1.3^\circ$ and the sensitivities at $2.0^\circ$ and $2.5^\circ$ are poor.
The significance to exclude NSI
is the best for the HK atmospheric
neutrino experiment and it is followed
by DUNE.  Notice that the sensitivity of
T2HK with the detector at Kamioka only
has poor sensitivity, and therefore
the second detector in Korea greatly
improves its sensitivity at all
the off-axis angles.
From the figure we notice that T2HKK at off-axis angle 1.3$^\circ$
can exclude the case with
($\epsilon_{ee}$, $|\epsilon_{e\tau}|$) = (0.8, 0.2)
at 2$\sigma$ within its proposed run-time for both the hierarchies. Whereas DUNE and HK can exclude the same by more than $3 \sigma$ in for NH. For IH the sensitivity of DUNE is similar that of the
$1.3^\circ$ configuration of T2HKK and the sensitivity of HK is around $2.5 \sigma$ in 15 years of running. The sensitivity of the T2HK experiment is less than $1 \sigma$ for both the hierarchies.

\subsection{CP violating phases
\label{cp}}
Next let us consider the implication
to the T2HKK experiment
in the case with an affirmative result of NSI.
As a reference value for NSI we take
($\bar{\epsilon}_{ee}$, $|\bar{\epsilon}_{e\tau}|$) = (0.8, 0.2),
which lies outside each exclusion curve
in the ($\epsilon_{ee}$, $|\epsilon_{e\tau}|$) plane
at 90\% C.L.\footnote{Notice that 
Fig.\,\ref{fig:fig2} is depicted for 3$\sigma$
and the allowed region at 90\% C.L., is smaller
than that at 3$\sigma$.}

\begin{figure*}
\hspace{-30pt}
\includegraphics[scale=1.1]{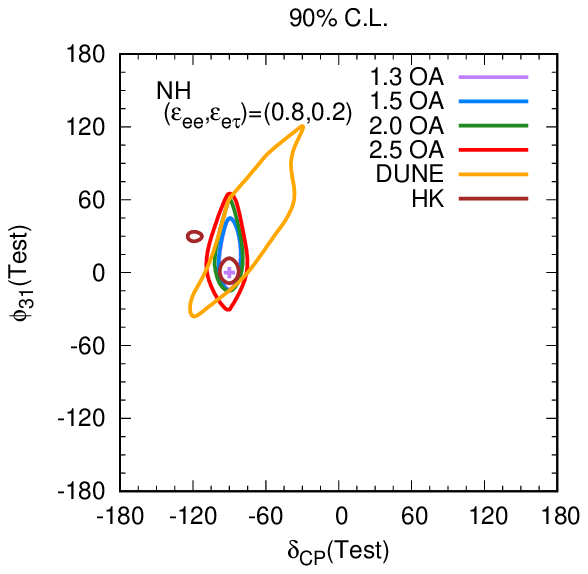}
\hspace{-90pt}
\includegraphics[scale=1.1]{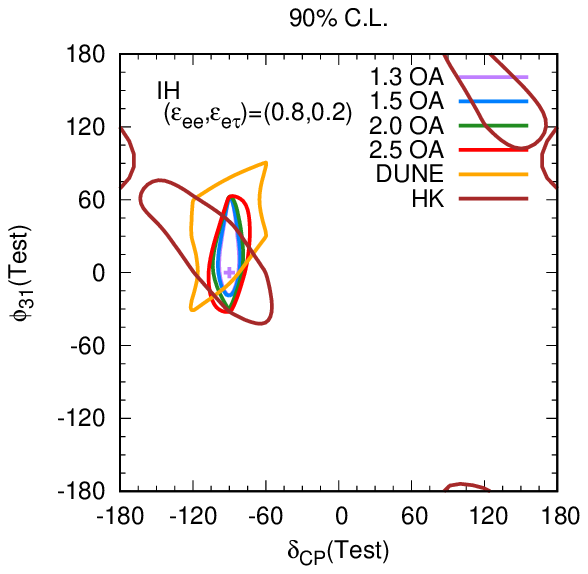}
\caption{The correlation between $\dcp$ and
$\phi_{31}\equiv$ arg($\epsilon_{e\tau}$) for
normal hierarchy (left panel) and inverted
hierarchy (right panel).  The true value
is ($\bar{\delta}_{\mbox{\rm\tiny CP}}$, $\bar{\phi}_{31}$) = ($-\pi/2$, 0).
} 
\label{fig:fig4}
\end{figure*}

\begin{figure*}
\hspace{-30pt}
\includegraphics[scale=1.1]{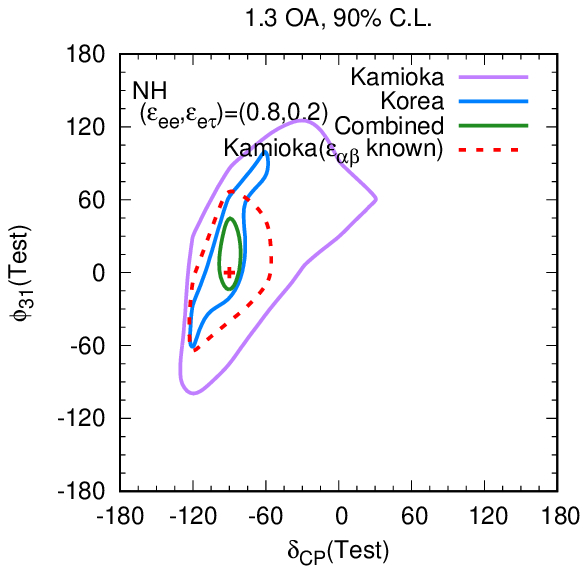}
\hspace{-90pt}
\includegraphics[scale=1.1]{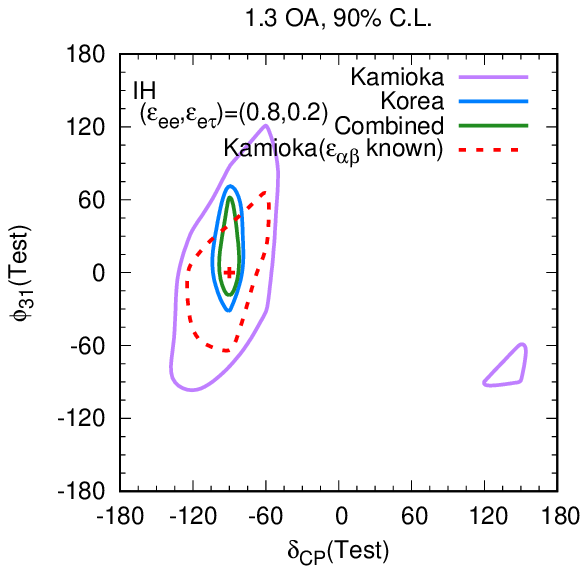}
\caption{The correlation between $\dcp$ and
$\phi_{31}\equiv$arg($\epsilon_{e\tau}$) for
normal hierarchy (left panel) and inverted
hierarchy (right panel) at the off-axis angle 1.3$^\circ$.
The true value are $\bar{\phi}_{31}=0$ and
$\bar{\delta}_{\mbox{\rm\tiny CP}}=-\pi/2$.
The dotted curves, which are given for
the detector at Kamioka without marginalizing over
$\epsilon_{ee}$ and $|\epsilon_{e\tau}|$,
are also shown to show the contribution
of these two parameters.
}
\label{fig:fig5}
\end{figure*}

The ansatz (\ref{ansatz}) contains the two
phases $\dcp$ and arg($\epsilon_{e\tau}$).
In the presence of NSI, it is important
how precisely we can determine these two phases.
So we study the correlation between $\dcp$ and arg($\epsilon_{e\tau}$)
around a certain set of the two phases.
Here we assume that the true oscillation
parameters are 
\begin{eqnarray}
\bar{\epsilon}_{ee} &=& 0.8\,,
\nonumber\\
|\bar{\epsilon}_{e\tau}| &=& 0.2\,,
\nonumber\\
\bar{\phi}_{31}&\equiv&\mbox{\rm arg}(\bar{\epsilon}_{e\tau})=0\,,
\nonumber\\
\bar{\delta}_{\mbox{\rm\tiny CP}}&=&-\frac{\pi}{2}\,.
\nonumber
\end{eqnarray}
The allowed regions at 90\% C.L., around the
true value ($\bar{\delta}_{\mbox{\rm\tiny CP}}$, arg($\bar{\epsilon}_{e\tau}$))
are shown in Fig.\,\ref{fig:fig4}
for ($\bar{\epsilon}_{ee}$, $|\bar{\epsilon}_{e\tau}|$) = (0.8, 0.2).
In these plots we have marginalized over $\epsilon_{ee}$ from $-4$ to $+4$ and $|\epsilon_{e \tau}|$ from 0 to 2.
To clarify the roles of the two detectors,
in the case of the off-axis angle 1.3$^\circ$,
separate contours are given in Fig.\,\ref{fig:fig5}
for the result from the detector in Kamioka (purple curve),
for that from the detector in Korea (blue curve), and for that from the combination of the two (green curve).
As we can see from Fig.\,\ref{fig:fig4},
T2HKK at the off-axis angles 1.3$^\circ$ and 1.5$^\circ$
has good sensitivity also in the sensitivity to the CP phases.
In the case of off-axis angle 1.3$^\circ$, the sensitivity
of T2HKK is better than that of DUNE.  This can be explained
as follows.
The detector at Kamioka, which has a shorter baseline length,
has poor sensitivity to the matter effect and therefore to
$\epsilon_{ee}$ and $|\epsilon_{e\tau}|$.
This is why the allowed region of the Kamioka detector
is large in Fig.\,\ref{fig:fig5} (purple contour) since the uncertainty in
$\epsilon_{ee}$ and $|\epsilon_{e\tau}|$ increases the uncertainty in the CP phases.  However, from the
result of the detector in Korea, we have stronger
constraint on $\epsilon_{ee}$ and $|\epsilon_{e\tau}|$.
If we use this information, then the detector at Kamioka
gives better sensitivity to $\dcp$ because of its
high statistics. To confirm this, in Fig. \ref{fig:fig5}, we also draw the contours for the Kamioka detector 
assuming $\epsilon_{ee}$ and $|\epsilon_{e\tau}|$ is known (the red dotted contours where we do not marginalize over $\epsilon_{ee}$ and $|\epsilon_{e \tau}|$) and we see that the allowed region shrinks profoundly.
So sensitivity of the combined T2HKK
detector complex to the CP phases is better than that
of DUNE.  This synergy of the detectors at Kamioka and in
Korea in the determination of the CP phases is the
striking advantage of the T2HKK experiment.
On the other hand, the HK atmospheric neutrino experiment
has disjoint allowed regions particularly in the
inverted mass hierarchy.  If one assumes that HK could
separate neutrinos and antineutrinos, then we have confirmed
that these disjoint regions disappear.
Thus as far as sensitivity to the CP phases is concerned,
its performance is not as good the T2HKK experiment
in inverted hierarchy. But for normal hierarchy the sensitivity of HK in constraining the CP phases is best among all the other setups because of the huge earth matter effects.

\section{Conclusion}
\label{conclusion}
We have studied the sensitivity of the T2HKK
experiment to the non-standard
interaction in propagation with the
ansatz (\ref{ansatz}).
With the ansatz (\ref{ansatz}), we obtained
the region in the ($\epsilon_{ee}$, $|\epsilon_{e\tau}|$)
plane in which T2HKK can distinguish NSI
from the standard three-flavor scenario.
As far as the sensitivity to NSI is concerned,
T2HKK at the off-axis angle 1.3$^\circ$ is the
best option, and with this option T2HKK
can discriminate NSI at 3$\sigma$ from
the standard case for approximately $|\epsilon_{ee}|\gtrsim 1$ and
$|\epsilon_{e\tau}|\gtrsim 0.2$. The sensitivity of DUNE is comparable as that of T2HKK with $1.3^\circ$ off-axis flux configuration in normal hierarchy but it is better in the inverted hierarchy. We find that
the sensitivity of the HK atmospheric experiment is the highest among the other setups considered in this work.

On the other hand, if the value of
$|\epsilon_{e\tau}|$ is relatively large
$|\epsilon_{e\tau}|\gtrsim 0.2$, then
we can determine the two phases 
$\dcp$, arg($\epsilon_{e\tau}$) separately
by T2HKK or DUNE.
As far as the sensitivity to the CP phases
is concerned, T2HKK is the better than DUNE.
The powerful feature of determination of
the two CP phases is the
remarkable advantage of the T2HKK experiment.
The atmospheric
neutrino experiment at HK is inferior to
the two long-baseline experiments in inverted
hierarchy but superior in normal hierarchy.

Since the matter effect $A$ and the baseline
length $L$ appears in the form of $AL/2\sim L/4000$ km
in the oscillation probability,
long-baseline neutrino experiments with longer baseline
lengths ($L\gtrsim$ 1000 km) are sensitive to the matter effect.
Hence they are also sensitive to NSI.
The nice feature of the T2HKK experiment is that
while the detector at Kamioka with a shorter baseline length
is advantageous to measure $\dcp$ because of its
high statistics, the one in Korea with a longer baseline
length has better sensitivity to the matter effect as well as
$\epsilon_{ee}$ and $|\epsilon_{e\tau}|$.
We have seen that the sensitivity to $\epsilon_{ee}$ and $|\epsilon_{e\tau}|$
is the best at the off-axis angle 1.3$^\circ$.
Thus we conclude that T2HKK at the off-axis angle 1.3$^\circ$ is
expected to be the best option to make the synergy of the
two detectors effective determine the NSI parameters
$\epsilon_{ee}$ and $|\epsilon_{e\tau}|$ as well as
the CP phases $\dcp$ and arg($\epsilon_{e\tau}$).

\section*{Acknowledgments}
The authors thank the Hyper-Kamiokande collaboration
for providing the neutrino flux of T2HKK in Korea. MG would like to thank Mark Hartz for many useful discussions and Chandan Gupta for help in ROOT.
This research was partly supported by a Grant-in-Aid for Scientific
Research of the Ministry of Education, Science and Culture, under
Grants No. 25105009, No. 15K05058, No. 25105001 and No. 15K21734.

\bibliography{t2hkkv02}

\end{document}